\documentclass{elsart}

\usepackage{graphicx}
                                                                                
\usepackage{amssymb}

\def\menorsim{\smash{\mathop{<}\limits_{\raise3pt\hbox{$\sim$}}}}
\def\maiorsim{\smash{\mathop{>}\limits_{\raise3pt\hbox{$\sim$}}}}
\def\xmax{X_{\rm max}}

\begin{document}

\begin{frontmatter}

\title{Strong colour fields and 
cosmic ray showers at 
ultra-high energies}

\author[sant]{J. Alvarez-Mu\~niz}
\author[centra,ist]{J. Dias de Deus}
\author[sant]{C. Pajares}

\address[sant]{Instituto Galego de F\'\i sica de Altas Enerx\'\i as \& 
Depto. de F\'\i sica de Part\'\i culas, \\
Univ. Santiago de Compostela, 15782 Santiago de Compostela, Spain}
\address[ist]{Departamento de F\'{\i}sica, IST, Av. Rovisco Pais, 1049-001 Lisboa, Portugal}
\address[centra]{CENTRA, Av. Rovisco Pais, 1049-001 Lisboa, Portugal}

\begin{abstract}
We argue that the increase of the ratio baryon/meson due to the presence
of strong colour fields and percolation in ultra-high energy hadronic 
collisions, helps to explain some of the global features of ultra-high energy 
cosmic ray cascades at $E>10^{18}$ eV and, in particular the observed
excess in the number of muons with respect to current models 
of hadronic interactions. A reasonable 
agreement with the small value and slope of the average depth 
of shower maximum $\langle \xmax \rangle$ vs shower energy -- as seen in data 
collected at the Pierre Auger Observatory -- can be obtained with
a fast increase of the p-Air production cross-section 
compatible with the Froissart bound.
\end{abstract}

\end{frontmatter}

The largest source of uncertainty in the prediction of observables
of cosmic-ray induced atmospheric showers at ultra-high energy (UHE) 
-- above $\sim 10^{17}-10^{18}$ eV -- stems from our limited knowledge
of the features of hadronic interactions in this energy range.
QCD-inspired models of multi-particle production need to be extrapolated to 
energies far beyond those reached in terrestrial accelerators,
and in regions of phase space of the collisions usually not covered 
in collider or accelerator experiments. On the other hand inferring 
the properties of the hadronic interactions at UHE in cosmic ray experiments 
is not an easy task which is hampered by the low luminosity 
of the primary beam, and by the unknown nature of its constituents.
Furthermore, data are relatively scarce at such high energies.

The composition of cosmic rays at high energies is still
a matter of controversy, mainly because direct measurements 
are only possible up to $E \sim 10^{15}$~eV~\cite{shibata}. 
Above this energy, 
attempts to infer the mass number $A$ of the primary particle   
are based on measured shower variables and are rather indirect~\cite{watson}.
Observables such as the number of muons at ground level $N_\mu$, or 
the depth at which the maximum in shower longitudinal development occurs $\xmax$,
are sensitive to both the nature of the primary and the hadronic interaction 
model \cite{watson&nagano}. In particular, the muon component of the 
shower is a powerful tool for the validation of hadronic interaction
models. 

Very recently an analysis of the data collected with the hybrid detector of the
Pierre Auger Observatory \cite{Augermuondata}, indicated that the number
of muons produced in UHE showers is about 1.5 times larger than that predicted 
by the QGSJETII \cite{QGSJETII} model assuming proton primaries. 
This conclusion seems to be 
rather independent of the primary cosmic ray composition. 
This ``deficit of muons" in the hadronic interaction models is  
consistent with the one found in the recent analysis
of more direct data on muons collected with the Yakutsk array \cite{Yakutskmuondata}.  

The Pierre Auger Collaboration has also reported  
that the arrival directions of 20 out of the 27 most energetic 
detected events above 57 EeV (1 EeV = $10^{18}$ eV) correlate  
with the positions in the sky of nearby Active Galactic Nuclei (AGN) \cite{AugerScience},
which are candidate sources of cosmic-ray production and acceleration
to ultra-high energies. It has been pointed out \cite{AugerScience} 
that this result together with 
the observed suppression of the cosmic-ray flux above
60 EeV \cite{Augerspectrum,HiResspectrum},  
is consistent with the hypothesis that most of the 
cosmic rays reaching the Earth are protons from nearby
astrophysical sources \cite{AugerScience}. 

There are also measurements of the average depth of maximum
shower development as a function of energy \cite{Augerxmaxdata,HiResxmaxdata}.  
The measured $\langle \xmax \rangle$ is smaller than what 
is predicted by current hadronic models assuming a proton dominated composition,
and moreover it increases with energy slower than expected \cite{Augerxmaxdata}.

In this work under the assumption of proton primaries, 
we show that the presence of strong colour fields in 
UHE collisions increases
the ratio of baryons to mesons ($B/M$) with respect to hadronic
models currently used in the simulation of atmospheric showers. 
This has the effect of increasing the number of muons, following 
the tendency seen in data. We also find that the behaviour of $\xmax$ is not strongly
affected by strong colour fields, and conclude that
available data can be explained with a rapid increase of the p-Air 
production cross section, but sill compatible with the Froissart bound.  

The global features of particle production in hadron-hadron (hh),
hadron-nucleus (hA) and nucleus-nucleus (AA) collisions seem
to be well described by models where particles are produced in the 
decay of longitudinal strings formed in the initial stage of 
the interaction 
\cite{Capellastrings,Ranftstrings,Wernerstrings,Amelinstrings,Nilssonstrings,Sorgestrings}. 
However, in order to explain some
effects observed at SPS and RHIC energies, such as the limited
growth of the rapidity particle density in the central region,
or the increase of the average transverse momentum and strangeness
with the number of participant nucleons \cite{RHIC}, 
one requires the introduction of fusion of strings \cite{Biro84,Amelin92} or 
more generally percolation of strings \cite{Armesto96,Pajares05}.
One should mention here that the results obtained within the string
percolation model coincide to a large extent with those
of the Colour Glass Condensate model \cite{McLerran,Lappi06}.

The key idea in the present work is that fusion and percolation
of strings lead to the formation of strings with higher string
tension, and this implies that mass effects -- formation of strange
quarks, di-quarks, etc... -- become less important. In the 
Schwinger model of string fragmentation \cite{Schwinger51}, the probability
of generating a set $S$ of quarks and anti-quarks and a conjugate
set $\bar S$ out of the vacuum, each one with mass $M$, is
\begin{equation}
P_M \sim \exp{\left(-\frac{\pi}{\kappa}~M^2\right)}
\label{eq:prob}
\end{equation}
where $\kappa$ is the string tension. As $\kappa$ tends to infinity
the probability (\ref{eq:prob}) becomes less and less dependent on $M$.
The role of strong colour fields to increase the fraction of heavier
particles has also been emphasized in \cite{Biro84,Amelin92,Amelin,Soff99,Topor05,Soff03}.

If we consider a single $q-\bar q~(3-\bar3)$ string, with $m$ being 
the mass of the u and d quarks ($m\simeq 0.23$ GeV), and $\kappa$ the 
3-representation string tension $\kappa\simeq 0.2~{\rm GeV}^2$, and neglect
strangeness and higher mass flavours, we have that the probability $B$ of 
producing a baryon is given by
\begin{equation}
B \sim \exp{\left(-\frac{\pi}{\kappa}~4m^2\right)}
\label{eq:probB}
\end{equation}
and the corresponding probability $M$ for meson production is
\begin{equation}
M \sim \exp{\left(-\frac{\pi}{\kappa}~m^2\right)},
\label{eq:probM}
\end{equation}
so that the ratio $B/M$ is
\begin{equation}
B/M \sim \exp{\left(-\frac{\pi}{\kappa}~3m^2\right)}\simeq 0.083.
\label{eq:ratioBM}
\end{equation}
This is our reference value for the ratio $B/M$ at low energies.

We shall next consider the effect of exchanging a quark by a diquark, 
as in Eqs.~(\ref{eq:probB}) and (\ref{eq:probM}), in more complex systems.
In $(q-\bar q)$ $(q-\bar q)$ fused strings, 
if they are in the SU(3) 6-representation with $\kappa_6=5\kappa/2$,
we obtain
\begin{equation}
\exp{\left(-\frac{\pi}{\kappa}~2m^2\right)}\simeq 0.19.
\label{eq:ratioBMfused6}
\end{equation}
which is larger than Eq.(\ref{eq:ratioBM}).
Furthermore, if we consider $(q-\bar q)$ $(q-\bar q)$ 
in the 8-representation with $\kappa_8=9\kappa/4$,
one obtains
\begin{equation}
\exp{\left(-\frac{\pi}{\kappa}~\frac{20}{9}m^2\right)}\simeq 0.16
\label{eq:ratioBMfused8}
\end{equation}
which is again larger than Eq.(\ref{eq:ratioBM}). 

In general, if we consider $N$ quarks in $(m,n)$ representations
of SU(3) (see for instance \cite{JeonSU3}), the highest dimension
representation corresponds to $N=m+2n$ with $m \simeq 2n \simeq N/2$,
and $\kappa_{(m,n)}\simeq \alpha\kappa N^2$ with $\alpha$ being 
a number of the order of $\sim 1/260$ such that for the quark
diquark interchange we obtain,
\begin{equation}
\frac{ \exp{(-\frac{\pi}{\kappa\alpha} \frac{(N+1)^2}{N^2}~m^2 )}}
     { \exp{(-\frac{\pi}{\kappa\alpha}~m^2})}
\label{eq:ratioBMfused}
\end{equation}
and, in the $N\rightarrow\infty$ limit,
\begin{equation}
\exp{\left(-\frac{\pi}{\kappa\alpha}~\frac{2m^2}{N}\right) \rightarrow 1}
\label{eq:largeNlimit}
\end{equation}
We take this result as an indication for the increase of the 
ratio $B/M$ as the energy or centrality of the collision 
increases.

It is also worth emphasizing that the enhancement of baryons or antibaryons over mesons is not only due 
to a mass effect as explained above. In fact in addition to the larger color 
and string tension in the percolation model, the way the cluster formed from 
the overlapping of individual strings decays, 
favours the increase of the ratio $B/M$ with energy 
and/or density of strings. 
Consider a cluster formed 
by several $q-\bar q$ strings. This cluster behaves as a $Q-\bar Q$ string, 
where $Q$ is composed of the different flavours of the individual $q-\bar q$ strings. 
The fragmentation of a cluster occurs through the successive creation of $Q-\bar Q$ 
complexes \cite{Amelin}.
Clearly this mechanism leads to an enhancement in the production of baryons over mesons, 
because the large number of quarks in the cluster of strings favours the formation of particles 
with higher number of constituents \cite{Pajares08}. 
Coalescence and recombination models \cite{coalescence} proposed similar mechanisms 
to explain the suppression of pion yields relative to baryons and antibaryons 
observed in relativistic heavy-nuclei collisions \cite{RHIC}. 

The increase with energy of $B/M$ helps to explain the excess
of muons in data when compared to existing hadronic interaction models. 
In fact, increasing the number of baryons and anti-baryons more muons are produced.
This effect has been shown with a simple toy model and also in 
the framework of the EPOS model of hadronic interaction in \cite{EPOSmuons}.   
Increasing the number of baryons decreases the number of 
neutral pions $\pi^0$ which decay into $\gamma \gamma$ 
and initiate electromagnetic subshowers in which muon production is typically very small.
Moreover a decrease in the amount of shower energy going into $\pi^0$s increases the total number of 
hadronic interactions in the shower, in which copious muon production typically occurs. 
Also $\pi^0$s tend to stretch shower development due to the production
of electromagnetic subshowers with comparatively larger $\xmax$ than 
hadronic subshowers of the same energy.  
As a consequence one would expect that decreasing the number of $\pi^0$s
-- increasing the $B/M$ ratio -- 
would limit the rate of increase of $\xmax$ with energy, a tendency 
seen in cosmic ray data. It is important to note that an increase
of the $B/M$ ratio also leads to a reduction in the number of charged pions, 
but as a first approximation charged pions play the same role as baryons 
in shower development and hence this has no consequences for either 
pion production or the behaviour of $\xmax$.   

In order to test these effects in a more quantitative way and 
make contact with Auger data, we have simulated sets of 100 proton
induced showers at energies ranging from $E=10^{15}$ eV up to $10^{20}$ eV.
The hybrid, one-dimensional shower simulation described in~\cite{jaimehybrid}
was used to obtain the average number of muons  at Auger ground 
level $\langle N_\mu \rangle$, and the average depth of shower 
maximum $\langle \xmax \rangle$. Firstly, we used the SIBYLL 2.1 hadronic
interaction model described in \cite{sibyll} which includes 
a ratio $B/M\sim 0.075$ constant with energy. Then we modified the 
SIBYLL 2.1 model and implemented  
a value of $B/M$ according to the following relation
\begin{equation}
B/M = 0.1 \log_{10}\frac{E}{\rm eV} - 1.3 
\label{BM}
\end{equation} 
normalized to the SIBYLL 2.1 value at low energy $E\sim 10^{14}$ eV
where accelerator data exists.

\begin{figure}[hbtp]
\begin{center}
\includegraphics[width=0.8\linewidth]{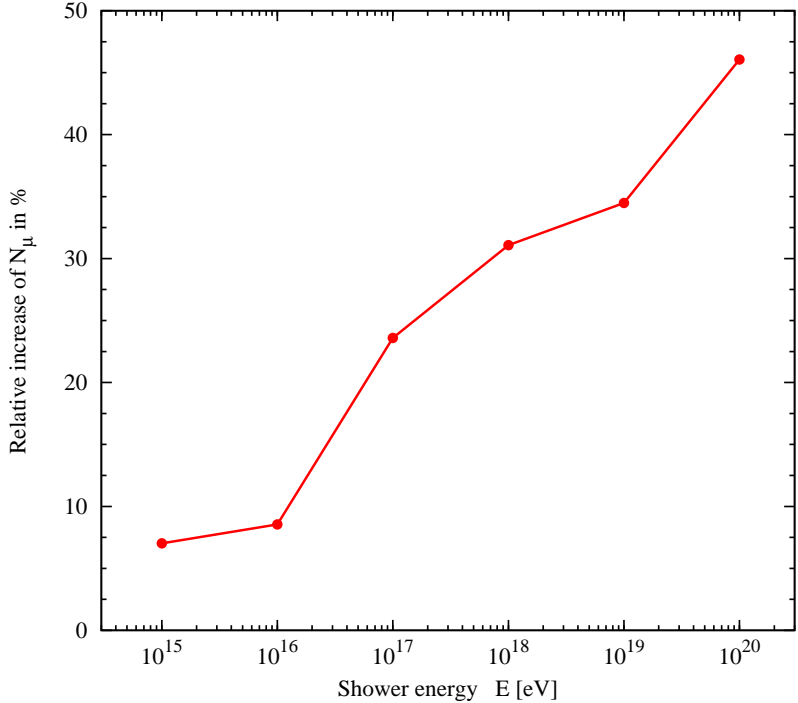}      
\end{center}
\caption{
Solid line: Relative difference between the average number of muons at Auger ground
obtained in proton-initiated showers simulated with the SIBYLL 2.1 
model, with a ratio $B/M\sim0.075$ constant with energy
$\langle N_\mu({\rm SIB})\rangle$, and with the SIBYLL 2.1 model but with 
a modified ratio $B/M$ following Eq.(\ref{BM}) $\langle N_\mu({\rm B/M})\rangle$.
The relative difference is calculated as 
$[\langle N_\mu({\rm B/M})\rangle-\langle N_\mu({\rm SIB})\rangle]/\langle N_\mu({\rm SIB})\rangle$ 
and is shown as a function of shower energy. The lines
joining the points are just to guide the eye.  
}
\label{fig:muons}
\end{figure}

In Fig.~\ref{fig:muons} we show the relative increase in the number of muons 
when increasing the ratio $B/M$. The relative difference between 
the average number of muons at ground predicted by SIBYLL 2.1 with $B/M\sim0.075$,
and the number of muons predicted by SIBYLL 2.1 but with 
$B/M$ following Eq.~(\ref{BM}) is shown as a function
of shower energy. One can see that at the highest energies 
the number of muons can increase
by as much as $50\%$ in agreement with Auger data \cite{Augermuondata}. 
It is important to remark that the choice of the underlying hadronic 
model is not relevant for our results, since we are only interested 
in the relative change of the number of muons when increasing the ratio $B/M$.

\begin{figure}[hbtp]
\begin{center}
\includegraphics[width=0.8\linewidth]{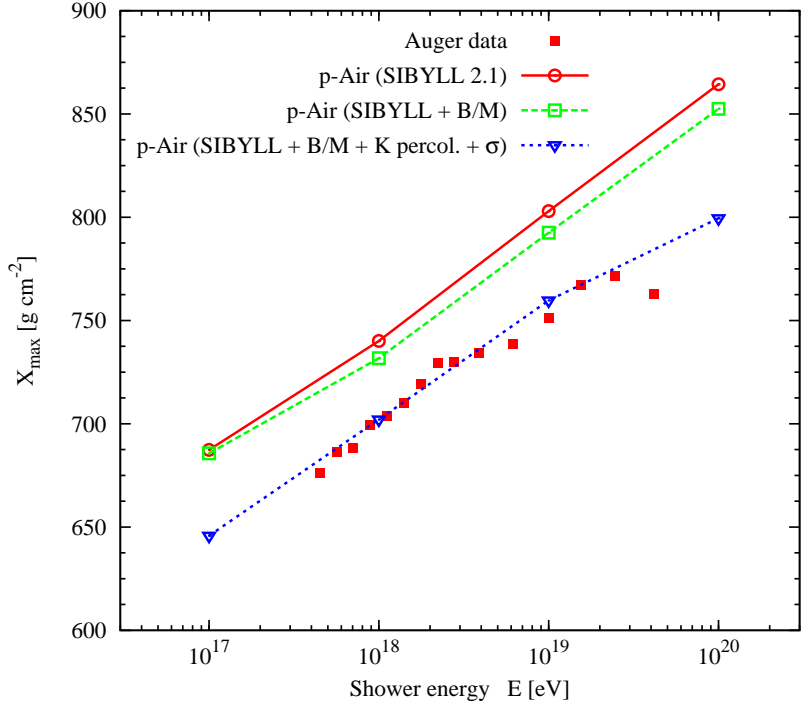}      
\end{center}
\caption{
Solid line with empty circles: Behaviour of the average depth of shower maximum 
$\langle\xmax\rangle$ with energy
as obtained in proton-initiated showers simulated with the SIBYLL 2.1 
model with ratio $B/M\sim0.075$ constant with energy.
Dashed line with empty squares: $\langle\xmax\rangle$ with the SIBYLL 2.1 model but with 
a modified ratio $B/M$ following Eq.(\ref{BM}).
Dotted line with empty circles: $\langle\xmax\rangle$ predicted   
by the SIBYLL 2.1 model but with 
a modified ratio $B/M$ following Eq.~(\ref{BM}), using the 
inelasticity $K$ predicted by the percolation model and with
the p-Air and $\pi-$Air cross section modified according
to Eq.~(\ref{xsection}). The filled squared points without lines are the data on $\langle\xmax\rangle$
collected at the Pierre Auger Observatory \cite{Augerxmaxdata}.
}
\label{fig:xmax}
\end{figure}

In Fig.~\ref{fig:xmax} we show the behaviour of $\xmax$ with 
energy as obtained in proton-induced showers with the SIBYLL 2.1
model, and with the SIBYLL 2.1 model with the ratio of $B/M$ modified
according to Eq.~(\ref{BM}). As explained above, increasing the ratio $B/M$ decreases the 
slope of $\langle\xmax\rangle$ vs $E$ curve and the value of $\langle\xmax\rangle$ 
due to the decrease in the 
production of $\pi^0$, which would otherwise induce electromagnetic
subshowers evolving deep into the atmosphere. One can see that the decrease 
of $\langle\xmax\rangle$ is not large enough to explain the trend 
observed in the data collected at the Pierre Auger Observatory \cite{Augerxmaxdata} 
also shown in Fig.~\ref{fig:xmax}.

Another important consequence of the presence of strong colour fields in 
the region where the interaction occurs is the percolation of strings.  
More details on the percolation model can be found in \cite{percolationbraun,percolationprl}. 
Of special importance for shower development is the prediction of 
the percolation model on the inelasticity $K$, defined as one minus the fraction of momentum 
carried by the fastest (leading) particle.  
Essentially, all existing high energy strong interaction models based on QCD 
and QCD evolution, predict an increase with energy of the 
inelasticity~\cite{qcd}. The same is true for the hadronic generators SIBYLL~\cite{sibyll} 
and QGSJET~\cite{QGSJETII}, widely used in the analysis of cosmic ray data.
However as shown in \cite{percolationprl}, above the 
percolation threshold the inelasticity decreases with energy.
This has important consequences for shower development as studied
in \cite{percolationprl,percolationjaime}, among them a decrease 
of the inelasticity tends to propagate the primary energy deeper 
into the atmosphere and increase $\xmax$ in contrast to what is seen in Auger data.
Note that the observed decrease of $\langle\xmax\rangle$ with energy \cite{Augerxmaxdata}
is at odds with the reported reduction of $A$ at high energies \cite{AugerScience}
As we shall see next, 
the behaviour of $\xmax$ with energy may be accommodated with 
an increase of the p-Air cross section compatible with the Froissart bound. 

\begin{figure}[hbtp]
\begin{center}
\includegraphics[width=0.8\linewidth]{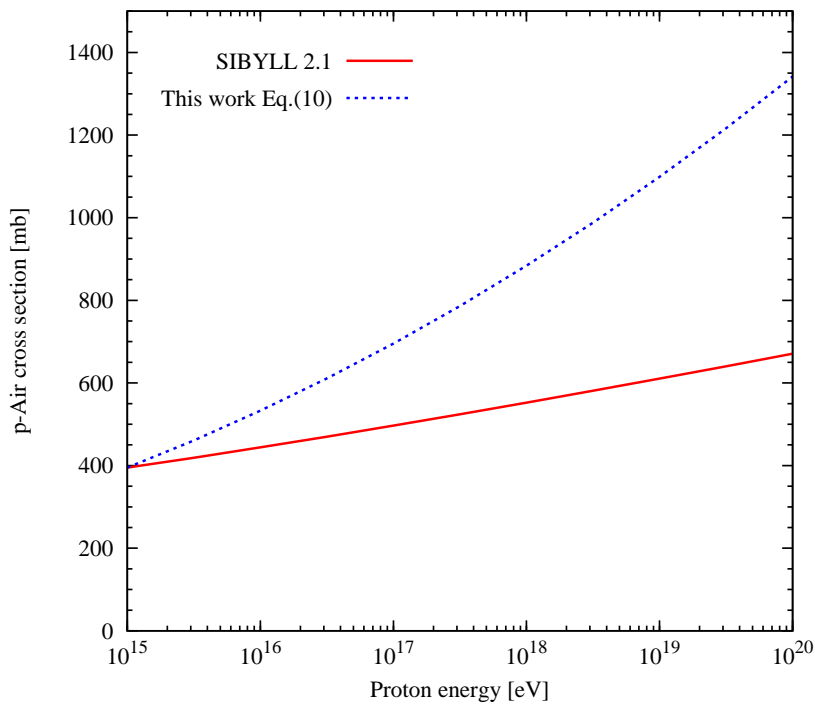}      
\end{center}
\caption{
The SIBYLL 2.1 p-Air cross section (solid line) compared
to the modified p-Air cross section (dashed line) in Eq.~(\ref{xsection}). 
Note that the SIBYLL cross section
behaves as $\log E $, while the modified cross section behaves
as $\log^2 E$, saturating the Froissart bound but not violating
it.}
\label{fig:xsection}
\end{figure}

We have implemented the inelasticity predicted by the percolation 
model \cite{percolationprl} in the SIBYLL 2.1 hadronic generator. 
Also we decreased the position of the first interaction and subsequent 
hadronic collisions by increasing the p-Air and $\pi$-Air production 
cross section in a similar manner as was done in \cite{Ulrichxsection}. 
In particular we have changed the SIBYLL 2.1 
energy dependence of the production cross section $\sigma_{\rm p-Air}^{\rm SIBYLL}$ 
using the relation
\begin{equation}
\sigma_{\rm p-Air} = \sigma_{\rm p-Air}^{\rm SIBYLL} 
(1 + 0.2\log_{10}\frac{E}{10^{15} {\rm eV}})
\label{xsection}
\end{equation}
which we apply for $E\geq 10^{15}$ eV, so that at lower energies  
one reproduces accelerator data. Moreover, since the SIBYLL cross section
behaves as $\log E $ the modified cross section in Eq.~(\ref{xsection}) clearly behaves
as $\log^2 E$, saturating the Froissart bound but not violating
it. The p-Air cross section in Eq.~(\ref{xsection}) and the SIBYLL 2.1 
p-Air cross section are shown in Fig.~\ref{fig:xsection}.

Our results for $\langle \xmax \rangle$ are shown in Fig.~\ref{fig:xmax} where 
a fair agreement with Auger data can be seen. 
This is mainly due to the changes implemented 
in the SIBYLL 2.1 cross-section, while modifying
the inelasticity and the $B/M$ ratio according to 
what strong colour fields induce, affects $\langle\xmax\rangle$ much less.
It is also interesting to note that an increase of the p-Air cross section also 
induces a decrease of the RMS of the $\xmax$ distribution with respect to 
what current models of hadronic interaction predict for proton-initiated showers.  

Also we have checked in our shower simulations, 
that the increase in the number of muons is fairly insensitive to 
both the decrease of inelasticity due to percolation of strings, and to the increase
of the p-Air and $\pi-$Air cross-section, the dominant effect
being the increase of the ratio $B/M$. 

We have shown that strong colour fields associated to high dimension representations of $Q-\bar Q$ strings, 
together with the $Q-\bar Q$ string breaking mechanism, lead to the increase of the $B/M$ ratio 
with respect to current hadronic models in which the $B/M$ ratio is $\sim 0.1$ and constant 
with energy. For UHE cosmic ray-induced showers this results  
in an increase of the average number of muons in the shower at ground level, 
and in a (small) decrease of $\langle\xmax\rangle$.
In order to further decrease $\langle\xmax\rangle$ 
and approximately agree with data, a large, but still compatible with the Froissart bound 
increase of the p-Air and $\pi$-Air cross sections with energy is required.

\section*{Acknowledgements}

We thank P. Brogueira, J.G. Milhano and M. Pimenta for helpful discussions on Auger data and strings.
J.~A-M. and C.~P. thank Ministerio de Educaci\'on y Ciencia
(FPA 2005-01963) and 
Spanish Consolider-Ingenio 2010 Programme CPAN (CSD2007-00042). 
J.~A-M. is also supported by the ``Ram\'on y Cajal" programme;  
by Ministerio de Educaci\'on y Ciencia (FPA 2007-65114);  
by Xunta de Galicia (2005 PXIC20604PN, PGIDIT 06 PXIB 206184 PR),
and by Feder Funds, Spain.
We thank CESGA (Centro de SuperComputaci\'on de Galicia) for computing resources.

\end{document}